\UseRawInputEncoding 

\documentclass[11pt]{article}%

\usepackage{amssymb}
\usepackage{amsmath}
\usepackage{graphics}
\usepackage{graphicx}
\usepackage{epsfig}
\usepackage{amsfonts}
\usepackage{graphicx}%

\usepackage{subfig}

\usepackage{booktabs}

\newcommand{\eqb}{\begin{equation}}
\newcommand{\eqe}{\end{equation}}
\newcommand{\dmb}{\begin{displaymath}}
\newcommand{\dme}{\end{displaymath}}

\newcommand{\eab}{\begin{eqnarray}}
\newcommand{\eae}{\end{eqnarray}}

\newcommand{\be}{\begin{equation}}
\newcommand{\ee}{\end{equation}}

\begin{document}
\begin{titlepage}
\begin{flushright}
\end{flushright}
\vspace{0.6cm}
\begin{center}
\huge{On emergent particles and stable neutral plasma balls in SU(2) Yang-Mills thermodynamics} 
\end{center}
\vspace{0.5cm}
\begin{center}
\large{Thierry Grandou$^{*}$ and Ralf Hofmann$^{**}$}
\end{center}
\vspace{0.1cm}
\begin{center}
{\em $^{*}$ Universit\'{e} Cote d'Azur\\ 
Institut de Physique de Nice, UMR CNS 7010\\ 
1361 routes des Lucioles, 06560 Valbonne, France;\\ 
thierry.grandou@inphyni.cnrs.fr 
}
\end{center}
\begin{center}
{\em $^{**}$ Institut f\"ur Theoretische Physik, Universit\"at Heidelberg,\\ 
Philosophenweg 12, D-69120 Heidelberg, Germany;\\ r.hofmann@thphys.uni-heidelberg.de
}
\end{center}
\vspace{1.5cm}
\begin{abstract}
For a pure SU(2) Yang-Mills theory in 4D we revisit the spatial (3D), ball-like region of radius $r_0$, in its bulk subject to the pressureless, deconfining phase at $T_0=1.32\,T_c$ where $T_c$ denotes the critical temperature for 
the onset of the deconfining-preconfining phase transition. Such a region possesses of a finite energy density and represents the self-intersection of a 
figure-eight shaped center-vortex loop if a BPS monopole of core radius $\sim \frac{r_0}{52.4}$, isolated from its antimonopole by repulsion externally invoked through a 
transient shift of (anti)caloron holonomy (pair creation), is trapped therein. The entire soliton (vortex line plus region of self-intersection of mass $m_0$ containing the monopole) 
can be considered an excitation of the pressureless and energyless ground state 
of the confining phase. Correcting an earlier estimate of $r_0$, we show that the vortex-loop  
self-intersection region associates with the {\sl central part} of a(n) (anti)caloron  
and that this region carries one unit of electric U(1) charge via the (electric-magnetic dually interpreted) 
charge of the monopole. The monopole core quantum vibrates at a thermodynamically determined frequency $\omega_0$ and is unresolved. 
For a deconfining-phase plasma oscillation about the zero-pressure 
background at $T=T_0$ we compute the lowest frequency 
$\Omega_0$ within a {\sl neutral} and homogeneous spatial ball (no trapped monopole) in dependence of its radius $R_0$. For 
$R_0=r_0$ a comparison of $\Omega_0$ with $\omega_0$ reveals that the neutral plasma oscillates much slower 
than the same plasma driven by the oscillation of a monopole core.

\end{abstract}
  
\end{titlepage}

\section{Introduction}

The thermodynamical phase structure of a single SU(2) Yang-Mills theory (electric-magnetic dually interpreted w.r.t. U(1)$\subset$SU(2)), comprising 
deconfining and preconfining thermal ground states of finite energy 
densities and (partially) massive gauge-field excitations as well as an energyless and pressureless 
confining ground state, suggests the existence of a solitonic, stable particle with intriguing yet familiar properties. 
Immersed into the confining ground state \cite{HofmannBook}, the figure-eight configuration of a self-intersecting center-vortex loop acquires 
its two-fold degenerate magnetic moment by a quantised electric current, composed of a chain of alternating 
monopoles and antimonopoles \cite{DelDebbio:1997}. 
The mass $m_0$ of this soliton mainly arises from the deconfining energy density within the 
self-intersection region of the vortex loop, idealised to be a {\sl spherical} blob of radius $r_0$. At a temperature $T_0=1.32\,T_c$, $T_c$ the boundary\footnote{For the sake of pointing out classical vs. quantum physics in association with (anti)calorons, 
we alternate between SI and natural units in this introduction, 
from Sec.\,\ref{RE} onwards we exclusively work in natural units where the speed of light in vacuum, 
Planck's (reduced) quantum of action, and Boltzmann's 
constant are all set equal to unity: $c=\hbar=k_B=1$.} of the deconfining phase, the 
pressure vanishes with positive slope (stable particle) \cite{HofmannBook}. 
The electric charge of this region is carried by a trapped BPS monopole of mass $m_m\ll m_0$ which is not 
resolved thermodynamically and orginated by large-holonomy (anti)caloron dissociation (pair creation) \cite{Diakonov2004}. 
The monopole's core fluctuates quantum initiated \cite{HofmannBook} to breathe at a certain frequency 
$\omega_0$ \cite{FodorRacz2004,Forgacs}.

It is tempting to interpret this quantum soliton as the electron (or an idealised charged lepton void of weak decay), 
see, however, \cite{Faber2001} for an interesting, extended field configuration, mapping Minkowski to the internal space $S_3$, argued to describe the electron's quantum numbers on the classical level in terms of topological charges. In the present paper, see also \cite{Hofmann2017}, we realise Louis de Broglie's ideas on 
the (quantum) thermodynamics of the isolated particle in its rest frame \cite{deBroglieDoc,DeBroglieRev} subject to 
internal, spatially homogeneous `clock ticks' at rate $\omega_0$. 
According to de Broglie, such a spatially localised (within a 3D ball-like volume), 
time-periodic phenomenon prescribes the soliton 
mass $m_0$ via $m_0 c^2=\hbar\omega_0$. Lorentz-boosting the standing wave of the particle at rest 
to a propagating wave (particle at spatial momentum $p$), this implies the well-known matter wavelength $\lambda=\frac{2\pi \hbar}{p}$ -- the starting point for the development of wave mechanics \cite{Schroedinger}. The 
probabilistic interpretation of the locus of electric charge in terms of the wavefunction's squared amplitude \cite{Born} 
matches with the fact that the periodically breathing, unresolved monopole core is displaced frequently and undeterministically 
by a local engagement with $\hbar$ (and an external field, e.g., in a hydrogen atom) throughout the self-intersection region of the vortex. 
The free solitons's apparent structurelessness, as inferred from collider experiments, is due to the thermal nature of the 
self-intersection region (maximum entropy): electric monopole charge equally likely occurs at any spatial point within the 
volume $\sim \frac{4}{3}\pi r_0^3$, in turn, immersed into the ground state of the 
confining phase: a condensate of shrunk-to-points center-vortex loops void of pressure and energy density \cite{HofmannBook}. (On the scale of rest mass $m_0$,  
the non-self-intersecting, spatial center-vortex loop is massless and prone to curve shrinking \cite{HofmannBook,MoosmannHofmannI}.)

In the present paper we discuss a few amendments to \cite{Hofmann2017}, where the 
above-sketched model for a charged lepton (idealised to be stable against weak decay) was introduced, and we address the oscillation 
physics about temperature $T_0$ of a neutral plasma ball to conclude that the 
contributions to the quantum mass of the soliton from the low-lying frequencies of such spherically symmetric breathing modes are negligible. 
In Sec.\,\label{RE} we correct the values of 
$r_0$ and the monopole mass $m_m$ as well as the ratio $m_m/m_0$ based on the monopole mass 
formula 
\eqb\label{massformulamonopole}
m_m=\frac{4\pi}{e(T_0)} H_\infty(T_0)\,
\eqe 
where $H_\infty(T_0)=\pi T_0$ \cite{HofmannBook,Diakonov2004,Nahm,KraanVanBaal,LeeLu}, and $e(T_0)=12.96$ is the value of the effective gauge coupling at 
$T_0$ \cite{HofmannBook}. In writing Eq.\,(\ref{massformulamonopole}) we assume that 
the spatial asymptotics of the monopole's $A_4$ (or adjoint Higgs) field is determined by temperature 
alone, reflecting the fact that the maximum holonomy of the originating, dissociated (anti)caloron is enforced externally and not influenced by 
the ensemble's spatial finite-range correlations encoded in the value of $e(T_0)=12.96\not=1$. 
The ratio $\frac{r_0}{|\phi|^{-1}(T_0)}$ is substantially smaller than erroneously 
estimated in \cite{Hofmann2017}: Instead of $\frac{r_0}{|\phi|^{-1}(T_0)}\sim 160$ we now have 
$\frac{r_0}{|\phi|^{-1}(T_0)}\sim 0.1033$. Here $|\phi|(T)=\sqrt{\frac{\Lambda^3}{2\pi T}}$ represents the modulus of the adjoint and 
{\sl inert} scalar field of the deconfining phase (spatially coarse-grained, densely packed   
(anti)caloron centers \cite{HofmannBook}), and $\Lambda$ denotes the Yang-Mills scale. 
The length scale $|\phi|^{-1}$ therefore sets the resolution for the SU(2) 
gauge-field theory prescribed by the quantum behavior of (anti)caloron centers \cite{HofmannBook}. 
In the present paper, we show that $\frac{r_0}{T_0^{-1}}\sim 1.29$. Therefore, the blob of center-vortex self-intersection\footnote{The radius $r_0$ is denoted $R_0$ in \cite{Hofmann2017}.} does not per se represent infinite-volume thermodynamics \cite{HofmannBook}. Rather, it is deeply contained within the 
central (quantum) region of the {\sl accomodating} caloron or anticaloron: $\frac{r_0}{|\phi|^{-1}(T_0)}\sim 0.1033$. 
We will argue, however, that, as far as the derivation of soliton properties is 
concerned, the thermodynamical results in \cite{HofmannBook} still apply. Interestingly, the smallness of $\frac{r_0}{|\phi|^{-1}(T_0)}$ 
excludes the possibility of trapping two or more monopoles or 
antimonopoles within one and the same blob: they would have to be provided by 
two or more dissociating (anti)calorons. In addition to a re-visit of the physics of the self-intersection region at temperature $T_0$ in a spatial center-vortex loop we also discuss in Sec.\,\ref{BreathingMode} the lowest radial oscillation of a {\sl neutral} deconfining plasma ball of radius $R_0$ 
which does not trap a monopole. Computing the associated frequency $\Omega_0$ requires the determination of the longitudinal sound speed $c_s$ at 
$T_0$. We obtain $c_s=0.479\,c$. Note that in spite of $T_0\sim T_c$, where conformal-symmetry breaking effects due to the Yang-Mills scale 
$\Lambda$ are large, this value is close to the ultra-relativistic-gas limit 
$c_s=\frac{1}{\sqrt{3}}\,c\sim 0.577\,c$. To exclude that there are sizable 
corrections to the left-hand side of Eq.\,(\ref{massfromm0}) due to plasma-breathing effects just above threshold we 
compare $\Omega_0$ and $\omega_0=m_0c^2/\hbar$ at $R_0=r_0$ and find $\Omega_0\ll\omega_0$. Finally, in Sec.\,\ref{Disc} 
we summarise our work and comment on future developments.

\section{Self-intersection region of a figure-eight shaped center-vortex loop\label{RE}}

In \cite{Hofmann2017} a model of the free electron was proposed which relies on the phase structure and thermodynamical quantities 
of SU(2) Yang-Mills thermodynamics, the nature of the excitations in the confining phase \cite{tHooft1978}, and 
the work in \cite{FodorRacz2004,Forgacs} on the perturbed BPS monopole. In \cite{tHooft1978} an operator -- the 't Hooft loop -- was defined nonlocally 
for pure SU(N) Yang-Mills theory. The 't Hooft loop is dual to the (spatial) Wilson loop. Its action creates one unit of magnetic flux w.r.t. 
the maximal Abelian subgroub U(1)$^{\rm N-1}\subset$SU(N) as expressed by a phase change through a 
root of unity in the Wilson loop linking to it. In the confining phase of SU(2) such a flux line occurs as the zero-core-size limit of the 
Abrikosov-Nielsen-Olesen vortex of winding number unity and thus is massless, causing the 't Hooft loop to acquire a 
finite, spatially homogeneous expectation. 
Moreover, no explicit isolated 
charges w.r.t. U(1)$\subset$SU(2)\footnote{In physics models, charges and fluxes w.r.t. U(1)$\subset$SU(2) need to be interpreted in an 
electric-magnetically dual way \cite{HofmannBook}.}, which could 
serve as flux sources or sinks, are tolerated by the pressureless confining 
ground state, composed of shrunk-to-points single center-vortex loops \cite{HofmannBook}. 
Therefore, a given, static vortex line, viewed as a 1D object in 3D space, has to form a closed loop. Yet, the self-intersection in a center-vortex loop locally represents a strong distortion 
of confining order (the 't Hooft loop representing the order parameter), re-instating a region of pressureless, deconfining phase wherein an isolated 
charge -- a BPS monopole or antimonopole -- may lead a shielded longtime existence. The response of such a classical BPS (anti)monopole 
to a spherically symmetric initial perturbation and the spectrum of normal modes were investigated 
in \cite{FodorRacz2004,Forgacs}, respectively. As a result, it was found that the asymptotic 
state of oscillation is determined by a frequency $\omega_0$ given by the mass of the two off-Cartan modes in the adjoint 
Higgs model that the monopole lives in. (In a thermal situation this adjoint Higgs model {\em is} 
the pure Yang-Mills theory with the $A_4$-component of the gauge field, whose value at spatial infinity determines 
the nontrivial (anti)caloron holonomy, playing the role of the Higgs field for the spatial components $A_i$ \cite{KraanVanBaal,LeeLu}.)   

We now revise the implications of the incorrect monopole mass formula in \cite{Hofmann2017} (Eq.\,(18) of \cite{Hofmann2017}). 
Also, we point out an error in Eq.\,(21) of \cite{Hofmann2017} which requires a conceptual re-interpretation 
of the physics of the self-intersection region. The mass $m_m$ of a BPS monopole is defined as the spatial integral of the 00-component of 
the energy-momentum tensor on its field configuration with winding number one in $\Pi_2({\rm SU(2)\setminus U(1)}=S_2)$ \cite{BPS}. It reads 
\eqb
\label{massBPSmon}
m_m=\frac{4\pi}{e}H_\infty\,.
\eqe
Here $e$ denotes the defining gauge coupling of the adjoint Higgs model (or the fundamental, thermalised pure SU(2) Yang-Mills theory), 
and $H_\infty$ is the spatially asymptotic 
modulus of the Higgs field. Only (anti)calorons of scale parameter close to $|\phi|^{-1}$ contribute 
to the emergence of the thermal ground state in the deconfining phase. Therefore, the gauge coupling $e$ can be interpreted 
as the effective one \cite{HofmannBook}. 

\begin{figure}
\centering
\includegraphics[width=15cm, height=6cm]{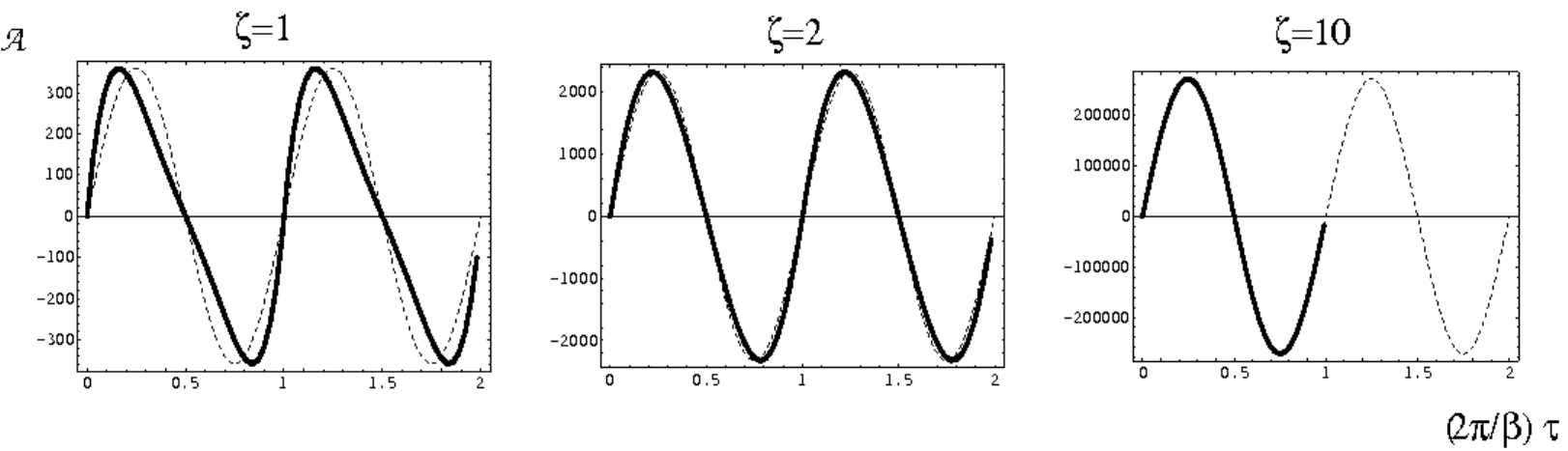}
\caption{Saturation towards a harmonic Euclidean time dependence of the contribution of a Harrington-Shepard 
caloron to the field-strength correlation defining the phase of the field $\phi$ as a function of the 
scaled cutoff $\xi\equiv \frac{\rho}{\beta}$ for the instanton-scale-parameter integration. 
Cutting off at $\rho\sim r_0=1.29\beta_0$ suggests that there are (mild) deviations from a harmonic dependence. 
Figure taken from \cite{HerbstHofmann2004}.\label{figsat}}
\end{figure}

Assuming that the monopole was liberated by the dissociation of a maximum-holonomy caloron 
at $T_0=1.32\,T_c$, we have \cite{HofmannBook,Diakonov2004}
\eqb
\label{eandH}
e(T_0)=12.96\,,\ \ \ \ \ H_\infty(T_0)=\pi T_0\,.
\eqe 
In writing the maximum nontrivial holonomy value of $H_\infty(T_0)$ we assume that this (externally imposed, pair creation) spatial asymptote solely 
depends on temperature and is not influenced by the spatial finite-range correlations imposed by the trivial-holonomy 
(anti)caloron constituting the thermal ground-state estimate. 
Being a quantum soliton of circular frequency $\omega_0$, the electron's rest mass $m_0$ is determined by the monopole-core 
oscillation \cite{deBroglieDoc,DeBroglieRev}. This frequency $\omega_0$ was found to be equal to the mass of the 
two off-Cartan modes in the adjoint Higgs model containing the BPS monopole \cite{FodorRacz2004,Forgacs}, 
\eqb
\label{coreoscfreq}
\omega_0=eH_\infty\,.
\eqe
On the other hand, assuming a conserved energy content of the approximately ball-like self-intersection region of radius $r_0$ (the contribution of the two vortex loops in negligible 
\cite{HofmannBook}, $r_0$ is denoted as $R_0$ in \cite{Hofmann2017}) -- trapping a quantum non-initialised, static BPS monopole of mass $m_m$ and constituted by a  
deconfining SU(2) Yang-Mills plasma of energy density $\rho(T_0)$ -- one derives the following equation \cite{Hofmann2017} 
\eab 
\label{massfromm0}
m_0&=&\omega_0=12.96\,H_\infty(T_0)=m_m+\frac{4\pi}{3}r_0^3\rho(T_0)\nonumber\\
&=&H_\infty(T_0)\left(\frac{4\pi}{12.96}+8.31\times\frac{128\pi}{3}\left(\frac{r_0}{18.31}\right)^3H_\infty^3(T_0)\right)\,.
\eae
Note that the left-hand side describes the soliton mass by a situation of an {\em initialised} monopole and its co-vibrating quantum surroundings -- all captured by a multiplication 
of the frequency $\omega_0$ with the quantum of action $\hbar$ {\sl after} the oscillation was triggered 
by monopole interaction with the thermal ground state and its excitations --, while the right-hand describes the energy balance of 
the system {\sl before} such an initialisation has occured. Energy conservation implies the equality of these two expressions. 
Notice that $m_0=12.96\,H_\infty(T_0)$ entails that the Yang-Mills scale $\Lambda$ relates 
to $m_0$ as \cite{Hofmann2017}
\eqb
\label{YMm0}
\Lambda=\frac{1}{118.6}\,m_0\,. 
\eqe
Solving Eq.\,(\ref{massfromm0}) for $r_0$ yields
\eqb
\label{R_0solv}
r_0=4.043\,H_\infty^{-1}
\eqe
instead of $r_0=4.10\,H_\infty^{-1}$ as obtained in \cite{Hofmann2017}. It is instructive 
to compute the relative contribution of $m_m$ to $m_0$, the ratio of the monopole core size $r_c$ to $r_0$, and the reduced Compton wave length $l_C$ to the thermodynamical resolution scale $|\phi|^{-1}(T_0)$. Note that $l_C=r_c=m_0^{-1}=\frac{1}{12.96\,H_\infty(T_0)}$ \cite{FodorRacz2004,Forgacs}. One has
\eqb
\label{relm_mtom_0}
\frac{m_m}{m_0}=\frac{4\pi}{(12.96)^2}=0.0748\,,\ \ 
\frac{r_c}{r_0}=\frac{1}{52.40}\,,\ \ 
\frac{l_C}{|\phi|^{-1}(T_0)}=0.00197\,.
\eqe 
Therefore, a hypothetically static, classical monopole would not contribute sizably to the mass of the region of 
self-intersection. Rather, it is the phenomenon of oscillation, expressing a strong interdependence between the unresolved monopole 
and its quantum environment, which gives rise to the mass of this region. The region's radius $r_0$ is about 50 times larger than 
the radius of the monopole core, and the monopole core intrinsically is far from being resolved. Therefore, it is safe to say that the quantum induced motion of the 
monopole is not influenced by the boundary. This boundary is the surface where plasma of the self-intersection region 
transmutes into a thin fuzzy/turbulent shell of preconfining phase adjacent to a surrounding of  
pressureless confining ground state which is composed of condensed single and shrunk-to-points center vortices \cite{HofmannBook}. Interestingly, 
radius $r_0$ compares to the spatial 
coarse-graining scale $|\phi|^{-1}(T_0)$ as follows
\eqb
\label{R0cgsVS}
\frac{r_0}{|\phi|^{-1}(T_0)}=\frac{4.043}{\sqrt{2\left(\frac{118.6}{12.96}\right)^3}}=0.1033\,.
\eqe   
A simple calculational error in Eq.\,(22) of \cite{Hofmann2017} has produced a value much larger than unity. 
The correct result of Eq.\,(\ref{R0cgsVS}) implies that the ``thermodynamics'' we discussed so far actually 
occurs deep within the center of a(n) (anti)caloron of scale parameter $\rho\sim |\phi|^{-1}(T_0)$. Since we have 
\eqb
\label{R0vsbeta}
\frac{r_0}{\beta_0}=1.29\ \ \ \ \ \ \ (\beta_0\equiv\frac{1}{T_0})\,,
\eqe
Fig.\,\ref{figsat} suggests that the scale parameter integral, defining the phase of 
$\phi$ \cite{HofmannBook}, does not quite saturate a harmonic dependence on Euclidean 
time when cut off at $\rho\sim r_0$. However, our above 
discussion on the right-hand side of Eq.\,(\ref{massfromm0}) assumes that such a saturation occurs
within the volume $\frac{4\pi}{3} r_0^3$. Hence the right-hand side of Eq.\,(\ref{massfromm0}) only yields an approximate account 
of the distorted thermodynamics within the self-intersection region: the monopole is always close 
to the locus of action at the inmost point of the caloron or anticaloron, rendering this 
region a jittery object even within its deep bulk.     
Still, the link between a hypothetically static monopole and its surroundings -- described by infinite-volume thermodynamics -- to the quantum-induced mass of 
the containing region due to oscillation is 
self-consistently made via Eq.\,(\ref{massfromm0}) by energy conservation, and reasonable 
estimates of $r_0$ and the critical temperature $T_c=13.87/(2\pi\times 118.6)\,m_0=9.49\,$keV \cite{Hofmann2017} ($\lambda_c\equiv 2\pi T_c/\Lambda=13.87$ 
\cite{HofmannBook}) should yet be possible.

\section{Lowest spherically symmetric breathing mode \label{BreathingMode}}

Spherically symmetric oscillations of the deconfining plasma temperature $T_0$ are not only driven 
by internally induced monopole oscillations but interfere with spatially homogeneous, coherent plasma oscillations due to external distortions of the initial 
temperature: $T_0\longrightarrow T_0+\delta T$. At the same radius $r_0$ these two oscillation modes 
should be compared in terms of their frequencies $\omega_0$ and $\Omega_0$, respectively, to judge whether the mass $m_0$ of the self-intersection 
region, arising from monopole-driven oscillation, is influenced sizably by the contribution of a global, oscillatory temperature distortion. 
Moreover, a neutral plasma ball of radius $R_0\gg r_0$ {\em is} subject to thermodynamics at face value. Since such a (distorted) macrocopic ball is 
expected to radiate electromagnetically on top of its weak black-body evaporation, diagnosing a lower frequency cutoff $\Omega_0$ in the excess 
spectrum would allow for the unique extraction of the plasma-ball radius $R_0$ and hence of the 
energy content of the ball.          

Let us thus compute the frequency $\Omega_0$ of 
the lowest spherically symmetric breathing mode of a deconfining SU(2) Yang-Mills 
plasma ball of homogeneous energy density $\rho$ and pressure $P$, whose temperature oscillates 
about $T_0$. Surface effects, arising from the transition between the 
deconfining (bulk) and the confining (exterior of ball) phases can be neglected 
for a sufficiently large ball mass $M\equiv\frac{4}{3}\pi R_0^3\rho(R_0)$, and 
the according expression for $\Omega_0$ reads \cite{Hartland2003,Lamb1882}
\eqb
\label{Omega}
\Omega_0=\frac{\pi c_s \Lambda}{\bar{R}_0}\,,
\eqe
where the dimensionless quantities $c_s$ (longitudinal sound velocity) and $\bar{R}_0$ ($R_0$ in units of the 
inverse Yang-Mills scale $\Lambda^{-1}$) are defined as
\eqb
\label{cs}
c^2_s\equiv\left.\frac{\frac{d\bar{P}}{d\lambda}}{\frac{d\bar{\rho}}{d\lambda}}\right|_{\lambda=\lambda_0}
\eqe
and $\bar{R}_0\equiv R_0\Lambda$. In Eq.\,(\ref{cs}) the employments of the one-loop pressure 
$P\equiv\Lambda^4\bar{P}$ and of the one-loop energy density 
$\rho\equiv\Lambda^4\bar{\rho}$ are excellent 
approximations (modified by higher-loop corrections on the 1\%-level \cite{HofmannBook}). Amusingly, an estimate of 
$\Omega_0$ by virtue of a linearisation of the force-balance equation 
$\frac{4}{3}\pi R^3\rho(R)\ddot{R}=4\pi R^2 P(R)$ about the stable point $R_0$ and an appeal to energy conservation, 
\eqb
\label{enercons}
\bar{R}\equiv R\Lambda=\left(\frac{3\bar{M}}{4\pi\bar{\rho}(\bar{R})}\right)^{1/3}\ \ \ \ \ \ \ 
(\bar{M}\equiv\Lambda M)\,,
\eqe
replaces the factor of $\pi$ in Eq.\,(\ref{Omega}) by a factor of three.  
For $\bar{P}$ and $\bar{\rho}$ we have \cite{HofmannBook}
\eab
\label{barrhoP}
\bar{P}(2a,\lambda)&\equiv&-\frac{2\lambda^4}{(2\pi)^6}\left[2\tilde{P}(0)+6\tilde{P}(2a)\right]-2\lambda\,,\nonumber\\ 
\bar{\rho}(2a,\lambda)&\equiv&\frac{2\lambda^4}{(2\pi)^6}\left[2\tilde{\rho}(0)+6\tilde{\rho}(2a)\right]+2\lambda\,,\nonumber\\ 
a&=&a(\lambda)\equiv 2\pi e(\lambda)\lambda^{-3/2}\ \ \ \ \ \ \ \ (e(\lambda_0)=12.96)\,,
\eae       
where 
\eab
\label{tilderhoP}
\tilde{P}(y)&\equiv&\int_0^\infty dx\,x^2\log\left[1-\exp\left(-\sqrt{x^2+y^2}\right)\right]\,,\nonumber\\ 
\tilde{\rho}(y)&\equiv&\int_0^\infty dx\,x^2\frac{\sqrt{x^2+y^2}}{\exp\left(\sqrt{x^2+y^2}\right)-1}\,.
\eae
Taking into account implicit (via $a(\lambda)$) and explicit dependences of 
$\bar{P}$ and $\bar{\rho}$ on $\lambda$ and employing the evolution 
equation for the mass of off-Cartan gauge modes as a function of temperature  \cite{HofmannBook}
\eqb
\label{alam}
1=-\frac{24\lambda^3}{(2\pi)^6}\left(\lambda\frac{da}{d\lambda}+a\right)a D(2a)\,,
\eqe
one derives 
\eab
\label{prhoder}
\frac{d\bar{P}}{d\lambda}&=&-
\frac{\lambda^3}{(2\pi)^6}\left(16\tilde{P}(0)+48(\tilde{P}(2a)-a^2D(2a))\right)\,,\nonumber\\ 
\frac{d\bar{\rho}}{d\lambda}&=&\frac{\lambda^3}{(2\pi)^6}\left(16\tilde{\rho}(0)+48\left(\tilde{\rho}(2a)- a^2(D(2a)-F(2a))\right)\right)+2\left(1-\frac{D(2a)-F(2a)}{D(2a)}\right)\,,\nonumber\\ 
\eae
where
\eab
D(y)&\equiv&\int_0^\infty dx\,\frac{x^2}{\sqrt{x^2+y^2}}\frac{1}{\exp\left(\sqrt{x^2+y^2}\right)-1}\,,\nonumber\\ 
F(y)&\equiv&\int_0^\infty dx\,x^2\frac{\exp\left(\sqrt{x^2+y^2}\right)}{\left(\exp\left(\sqrt{x^2+y^2}\right)-1\right)^2}\,.
\eae
Substituting Eqs.(\ref{prhoder}) into Eq.\,(\ref{cs}) at $\lambda_0=18.31$, we numerically obtain 
\eqb
\label{csnum}
c_s(\lambda_0)=0.479\,.
\eqe
For Eq.\,(\ref{Omega}) this yields 
\eqb
\label{om0num}
\Omega_0=1.506\,\frac{\Lambda}{\bar{R}_0}\,.
\eqe
Let us now compare the monopole-core induced frequency $\omega_0$ of the self-intersection 
region of the figure-eight shaped center-vortex loop (model of the electron) 
with $\Omega_0$ at one and the same radius 
\eqb
\label{radiuseq}
r_0=R_0=4.043\,H_\infty^{-1}(T_0)\,,
\eqe
see Eq.\,(\ref{R_0solv}). For $\Omega_0$ this yields
\eqb
\label{omegar0}
\Omega_0=0.372\,H_\infty(T_0)
\eqe
such that 
\eqb
\label{ratio_frequ}
\frac{\omega_0}{\Omega_0}=\frac{12.96}{0.372}=34.84\,.
\eqe
Such a large ratio is natural since the 
oscillation in the self-intersection region -- quantum initiated by caloron or anticaloron action -- is induced by the classical 
dynamics of a monopole core \cite{FodorRacz2004,Forgacs} whose radius matches the 
reduced Compton wave length $l_C$ while the lowest symmetric breathing mode of the neutral 
deconfining ball is a consequence of longitudinal sound propagation in an approximate bulk thermodynamics. This bulk 
associates with $r_0$ being comparable to the Bohr radius \cite{Hofmann2017}. 

Eq.\,(\ref{om0num}) is the more reliable the larger $\bar{R}_0$ is. Isotropy breaking effects, which associate with the neglected surface dynamics of the ball and/or the excitation of spherically non-symmetric oscillation states, cause this surface to (electromagnetically) radiate with a spectrum that is cut off towards the infrared at a frequency of about $\nu_0\sim \frac{\Omega_0}{2\pi}$, corresponding to a wave length 
$l_0=\frac{1}{\nu_0}\sim \frac{2\pi R_0}{1.506}$.

\section{Summary and Discussion \label{Disc}}

This paper's main purpose was to compare two situations in which a ball-like 
region of deconfining phase in SU(2) Yang-Mills thermodynamics 
oscillates about the zero of the pressure at temperature $T_0$: 
(i) the charged self-intersection region of a figure-eight shaped, solitonic 
center-vortex loop (a model of the electron) containing an internally quantum-perturbed BPS 
monopole, whose classical core dynamics drives this oscillation of (circular) frequency $\omega_0$ 
(up to a factor $\hbar$ coincident with the rest energy $m_0c^2$ of the soliton \cite{deBroglieDoc,DeBroglieRev}), and 
(ii) the homogeneous, neutral region being perturbed externally such that a lowest spherically symmetric oscillatory 
excitation of (circular) frequency $\Omega_0$ is excited thanks to a finite longitudinal speed of sound $c_s$. At the same 
radius, $r_0=R_0=\frac{4.043}{\pi T_0}$, we obtain $\frac{\omega_0}{\Omega_0}=34.84$. This hierarchy relates to the very different 
causes of oscillation in either case and assures that the right-hand side of the 
mass formula Eq.\,(\ref{massfromm0}) for an oscillation in the sense of (i) 
does not receive any sizable contributions from an oscillation in the sense of (ii).  

Secondly, we have noticed a numerical error in \cite{Hofmann2017} concerning the estimate of the system radius $r_0$ in terms of the spatial coarse-graining 
scale $|\phi|^{-1}$ at $T_0$. The correct result 
states that finite-size corrections to infinite-volume thermodynamics cannot be excluded at face value since the region of 
self-intersection actually is contained deeply within the ball of 
spatial coarse-graining invoked in the derivation of the effective theory \cite{HofmannBook}. 
However, the asymptotic harmonic time dependence of the integrated field-strength correlation \cite{HofmannBook}, required for the 
introduction of the field $\phi$, is approximately saturated when cutting the instanton-scale-parameter integration 
off at $r_0<|\phi|^{-1}$ already.  

The present work only represents a first step in studying the plasma 
dynamics of a ball-like region of deconfining phase at $T_0$. More 
realistically, the physics of the boundary shell, repesenting the transitions from the deconfining via the preconfining to the confining phases,   
should be taken into account. Also, we did not 
address the evaporation physics (in adiabatic approximation: emission of non-intersecting and self-intersecting 
center-vortex loops and, assuming a mixing with an SU(2) Yang-Mills theory of much lower scale, 
electromagnetic modes from the surface of this shell in terms of thermal spectra) in case of macroscopically 
sized balls, see \cite{GH2009}, and how this process 
affects the oscillation dynamics. Last but not least, a thorough discussion of (delocalised) spin in terms of the center-flux along the figure-eight 
shaped vortex line and the emergence of the electric Coulomb field \cite{Faber2022} throughout the confining-phase exterior 
to the blob of vortex-line self-intersection needs to be understood. The latter represents a small contribution to the 
soliton mass \cite{Hofmann2017} and should manifest polarised dipole densities in (anti)caloron peripheries \cite{HofmannBook,GrandouHofmann2015}. 

Our results on the spherically symmetric oscillations 
of the homogeneous and macroscopic plasma could be relevant in the 
description of certain, quasi-stabilised, compact and radiating objects created 
within atmospheric discharges and for plasma diagnostics in terrestial fusion experiments. \vspace{0.1cm}\\

\noindent\textbf{Acknowledgments:} We would like to acknowledge a useful 
conversation with Anton Plech. Discussions with Manfried Faber on common themes and differences of his and 
the here-proposed soliton model are gratefully acknowledged. \vspace{0.1cm}\\

\noindent\textbf{Author contributions:} The material of Secs.\,2 and 3 was originally 
conceived by RH. Both authors, TG and RH, have contributed
equally to an intense discussion, correction, and 
conceptual consolidation of this first step. \vspace{0.1cm}\\ 

\noindent\textbf{Funding:} None. \vspace{0.1cm}\\  

\noindent\textbf{Conflicts of Interest:} There is no conflict of interest.






\end{document}